\documentclass[prb,reprint,letterpaper,noeprint,longbibliography,nodoi,footinbib]{revtex4-1} 

\usepackage{hyperref}
\hypersetup{colorlinks, allcolors=blue}
\usepackage{amsmath}  
\usepackage{amsfonts} 
\usepackage{graphicx} 

\newcommand{\del}[2]{\frac{\partial{#1}}{\partial{#2}}}
\newcommand{\lag}{\mathcal{L}}

\begin{document}

\title{External Potentials and Ehrenfest Relations in\\ Lagrangian Field Theories}

\author{Rayn Samson}
\email{samsonrayn@grinnell.edu} 

\affiliation{Department of Physics, Grinnell College, Grinnell, Iowa 50112, USA}

\date{\today}

\begin{abstract}

This paper develops a general method to construct Ehrenfest-like relations for Lagrangian field theories when an external, coordinate-dependent scalar potential is applied. To do so, we derive continuity equations in which the spatial and temporal derivatives of the potential can be interpreted as a source of field momentum and field energy, respectively. For a non-relativistic Schr\"{o}dinger field theory, these continuity equations yield Ehrenfest's theorem for energy, linear momentum, and angular momentum. We then derive a relativistic counterpart for these relations using complex Klein-Gordon fields coupled with an electric potential.  

\end{abstract}

\maketitle

\section{Introduction}
It is well understood in classical mechanics that the conservation of linear momentum in a system only holds when no external force is acting on it. Put in the language of Noether's theorem, a system is not symmetric under translation when a spatially dependent potential acts on it---in which case the conservation of linear momentum is broken. However, much of the richness in classical mechanics comes from understanding how mechanical momentum changes in time when such a potential is applied. This is captured by Newton's second law,
\begin{equation}
    \frac{d}{dt}\mathbf{p}=-\nabla V.
\end{equation}
Of course, this expression reduces to a statement of momentum conservation when the potential $V$ is not spatially dependent. While this result can be easily derived as the equation of motion in a classical Lagrangian of the form $L(\mathbf{q},\dot{\mathbf{q}},t)$, there has so far been no systematic way to produce analogous results for the field momentum of a field Lagrangian $\lag(\varphi_1,\dots, \varphi_n,\partial_\mu \varphi_1,\dots,\partial_\mu \varphi_n)$. This paper develops such a method by first considering a non-relativistic field theory that describes the Schr\"{o}dinger equation with an external, coordinate-dependent potential in Section \ref{Schrodinger Field Theory}. Conservation of energy and momentum does not hold in general for such a system, but a close examination of the potential's response to coordinate transformations yields Ehrenfest's theorem for linear and angular momentum as well as a similar result for energy. We then use a generalized version of the same method in Section \ref{KG} to derive analogous results in a complex Klein-Gordon field theory with minimal electric potential coupling. In this paper, such results are referred to as ``Ehrenfest" or ``Ehrenfest-like" relations owing to their structural similarities with the original results obtained from the Schr\"{o}dinger field theory. However, the content of relativistic Ehrenfest relations has an entirely different interpretation from the non-relativistic case.    

    \subsection{Notation and Conventions}
    We will be working in flat Minkowski spacetime with Cartesian coordinates. The mostly positive signature $(-+++)$ is used. Coordinate derivatives are abbreviated with a comma before the appropriate index, e.g. $ M_{,\mu}\equiv \partial_\mu M$. Greek indices run through all four spacetime coordinates, while latin indices run through the three spatial coordinates. We will switch freely between spatial index notation and vector notation. For example, the four-divergence is given by 
    \begin{equation}
        \partial_\mu\equiv\begin{pmatrix}
            \partial_0\\
            \nabla
        \end{pmatrix},
    \end{equation}
    where $\nabla=(\partial_1,\partial_2,\partial_3)$ may be represented with the open spatial index operator $\partial_i$.

\section{Noether Currents}
    In the context of classical field theory, Noether's theorem describes a fascinating equivalence between continuous symmetries of a Lagrangian system and its conserved quantities. Starting with an action of the form
    \begin{equation}\label{eq:action}
        \mathcal{S}=\int \lag(\varphi,\partial_\mu\varphi)\,d^4x,
    \end{equation}
     where $\lag$ is a Lagrangian density (henceforth: ``Lagrangian") that depends on a field $\varphi$ and its derivatives. Using the principle of stationary action, the field $\varphi$ must satisfy the Euler-Lagrange field equation of motion
     \begin{equation}\label{eq:fieldeq}
         -\partial_\mu \left(\del{\lag}{\varphi_{,\mu}} \right)+\del{\lag}{\varphi}=0.
     \end{equation}
    A ``symmetry" of the system is a transformation that yields the same physical predictions. A transformation of the field,
     \begin{equation}\label{eq:transformation}
     \varphi \to \varphi +\delta\varphi,
     \end{equation}
     is a symmetry if it leaves Eq. (\ref{eq:fieldeq}) invariant. This holds most generally when the transformation deforms the Lagrangian by a divergence of some function $K^\mu$:
     \begin{equation}\label{eq:surface}        
     \lag\Big(\varphi+\delta\varphi,\partial_\mu(\varphi+\delta \varphi)\Big)-\lag(\varphi,\partial_\mu \varphi )=\partial_\mu K^\mu.
    \end{equation}
     (We will later argue that the reverse is not true: invariance of Eq. (\ref{eq:fieldeq}) does not imply symmetry under a given transformation.) Calculating the left-hand side of Eq. (\ref{eq:surface}) with a series expansion and simplifying with the equation of motion yields
     \begin{equation}
         \lag\Big(\varphi+\delta\varphi,\partial_\mu(\varphi+\delta \varphi)\Big)-\lag(\varphi,\partial_\mu \varphi ) = \partial_\mu\left(\del{\lag }{\varphi_{,\mu}}\delta \varphi\right).
     \end{equation}
     Equating this to the surface term $\partial_\mu K^\mu$ yields the vanishing divergence
    \begin{equation}\label{eq:conservation}
        \partial_\mu\left(\del{\lag }{\varphi_{,\mu}}\delta \varphi-K^\mu\right)=0,
    \end{equation}
    which defines a conservation law in the form of a continuity equation. The term in parentheses defines a \textit{Noether current}. Importantly, if the action depends on more than one field, we would need to add the response of each field to the transformation when defining the Noether current. In this paper, we will examine Lagrangian systems that depend on two fields which are complex conjugates of each other, so Noether currents will take the form
    \begin{equation}\label{eq:noether_current}
        J^\mu\equiv \del{\lag}{\varphi_{,\mu}}\delta \varphi+\del{\lag}{\overline{\varphi}_{,\mu}}\delta \overline{\varphi} -K^\mu.
    \end{equation}
    We now proceed to discuss how the conservation of these Noether currents is affected by an external scalar potential.
    
    \section{Schr\"{o}dinger Field Theory}\label{Schrodinger Field Theory}
    Suppose we are given a physical system that is influenced by an external, coordinate-dependent potential. One such example is a system described by the Schr\"{o}dinger equation
    \begin{equation}\label{eq:schrodeq}
        -\frac{\hbar^2}{2m}\nabla^2\Psi + V\Psi=i\hbar \del{\Psi}{t}.
    \end{equation}
    Crucially, Eq. (\ref{eq:schrodeq}) does not specify the source or dynamics of the potential $V$, which is why we say that it is \textit{external} to the theory. To formulate this equation in terms of field theory, consider a \textit{Schr\"{o}dinger Lagrangian}\cite{Franklin2017}\cite{TakahashiRop1986} 
    \begin{equation}\label{eq:shrodlag}
    	    \lag=\frac{\hbar^2}{2m}\partial_j \overline{\Psi} \partial^j \Psi +V\overline{\Psi}\Psi-\frac{i\hbar}{2}\left(\overline{\Psi}\dot{\Psi}-\Psi\dot{\overline{\Psi}}\right),
    	\end{equation}
    where the fields are $\Psi$ and $\overline{\Psi}$. The equation of motion associated with the latter,
    \begin{equation}
    	    -\del{}{t}\left(\del{\lag}{\dot{\overline{\Psi}}} \right)-\del{}{x^j}\left(\del{\lag}{\overline{\Psi}_{,j}} \right)+\del{\lag}{\overline{\Psi}}=0,
    \end{equation}
    reduces to Eq. (\ref{eq:schrodeq}). To account for the potential's response to a coordinate transformation, we treat the Lagrangian as an explicit function of $V$ in addition to the fields and their derivatives:
    \begin{equation}\label{eq:dependence}
        \lag = \lag(\Psi,\overline{\Psi}, \partial_\mu\Psi, \partial_\mu\overline{\Psi}, V).
    \end{equation}
    Despite the awkward dependence on the potential, this Lagrangian still admits a $U(1)$ symmetry. Consider the complex rotation $\Psi\to e^{i\theta}\Psi$ (and cc).
    For small $\theta$ we have $e^{i\theta}\approx 1+i\theta$, and so the fields respond with
    \begin{equation}
        \Psi \to \Psi + i\theta\Psi \quad\text{and}\quad  \overline{\Psi}\to \overline{\Psi}-i\theta\overline{\Psi}.
    \end{equation}
    The associated conservation law becomes a statement of probability conservation,
        \begin{equation}\label{eq:probcons}
             \del{}{t}|\Psi|^2+\nabla \cdot \mathbf{j}=0,
        \end{equation}
    where $|\Psi|^2\equiv \overline{\Psi}\Psi$ is interpreted as a probability density and $\mathbf{j}$ is the probability current defined by\cite{Griffiths2018-vs}
    \begin{equation}
    \mathbf{j}\equiv \frac{\hbar}{2m i}\left(\overline{\Psi}\nabla \Psi-\Psi\nabla\overline{\Psi}\right).
    \end{equation}
    This result is unaffected by the presence of $V$ in the Lagrangian, but the same is not true for coordinate transformations. Consider a spacetime translation 
    \begin{equation}\label{eq:spacetime translation}
    x^\mu\to x^\mu +a^\mu,    
    \end{equation}
     for some (small) constant $a^\mu$. The fields are scalars, so they transform as
    \begin{equation}
        \Psi \to \Psi + a^\mu\partial_\mu \Psi \quad \text{and} \quad \overline{\Psi}\to \overline{\Psi}+ a^\mu \partial_\mu \overline{\Psi},
    \end{equation}    
    and crucially, so does the potential: 
    \begin{equation}
        V \to V+a^\mu \partial_\mu V.
    \end{equation}
    As a scalar, the Lagrangian also transforms as $\lag \to \lag + a^\mu \partial_\mu \lag$. Thus the change in the Lagrangian (which we will denote as $\delta \lag$) calculated by treating $\lag$ as a scalar is given by
    \begin{equation}\label{eq:deltalag_scalar}
        \delta \lag = \partial_\mu(a^\mu\lag),
    \end{equation}
    which gives the surface term. Care must be taken to understand what is meant here: as a scalar, the Lagrangian does changes by an overall divergence, and so the equations of motion are invariant even in the presence of $V$. However, computing $\delta \lag$ from a first-order series expansion in all arguments of the transformed Lagrangian does not yield an overall divergence, and dependence on $V$ is the culprit: 
    \begin{equation}\label{eq:deltalag_taylor}
        \delta \lag = a^\nu\left[\partial_\mu\left(\del{\lag}{\Psi_{,\mu}}\Psi_{,\nu}+\del{\lag}{\overline{\Psi}_{,\mu}}\overline{\Psi}_{,\nu} \right)+ \del{\lag}{V}\partial_\nu V\right]. 
    \end{equation}
   We can thus see that $\lag$ is not translationally symmetric, and hence there is no conservation law to be obtained from a vanishing divergence. That is, invariance of a potential-dependent Lagrangian under a coordinate transformation \textit{does not} imply a symmetry under that transformation. However, we can still set Eqs. (\ref{eq:deltalag_scalar}) and (\ref{eq:deltalag_taylor}) equal to each other and see what is obtained instead of a conservation law. After discarding the overall constant $a^\nu$, we define the energy-momentum tensor in a standard way:
   \begin{equation}\label{eq:schrod stress tensor}
    	 T^\mu_{\,\,\,\nu}\equiv \del{\lag}{\Psi_{,\mu}}\Psi_{,\nu}+\del{\lag}{\overline{\Psi}_{,\mu}}\overline{\Psi}_{,\nu} -\delta^\mu_{\,\,\,\nu} \lag.
    \end{equation}
    We still say that $T^\mu_{\,\,\,\nu}$ represents four Noether currents containing the energy ($\nu=0$) and momentum ($\nu=1,2,3$) densities stored in the fields. However, the divergence of $T^\mu_{\,\,\,\nu}$ is not in general zero. Instead, the derivatives of the potential act as a ``source" in the continuity equations
   \begin{equation}\label{eq:source eq}
       \partial_\mu T^\mu_{\,\,\,\nu}= -\del{\lag}{V}\partial_\nu V.
   \end{equation}
     The conservation of these currents depends on the presence of an external potential and its derivatives. In analogy with classical mechanics, the momentum of a system is conserved only when there is no coordinate-dependent external potential acting on it. When an external potential is applied, momentum is no longer conserved and the change in momentum is summarized by Newton's second law: $d\mathbf{p}/dt=-\nabla V$. The equivalent result in non-relativistic quantum mechanics would be Ehrenfest's theorem for momentum, $d \langle \hat{\mathbf{p}}\rangle/d t =\langle -\nabla V\rangle $. Section \ref{ehrenfest relations} will show that this is one of two statements contained in Eq. (\ref{eq:source eq}).
     
    \subsection{Ehrenfest Relations} \label{ehrenfest relations}
    Having defined the energy-momentum tensor associated with the Schr\"{o}dinger Lagrangian, it is helpful to write out the components explicitly:
    \begin{align}
    	T^0_{\,\,\,0}&=-\frac{\hbar^2}{2m}\partial^i \overline{\Psi} \partial_i \Psi -V\overline{\Psi}\Psi         \label{eq:SC-00} \\
    	T^i_{\,\,\,0}&=\frac{\hbar^2}{2mc}\left(\dot{\overline{\Psi}}\partial^i\Psi+\dot{\Psi}\partial^i\overline{\Psi} \right) \label{eq:SC-i0} \\
            T^0_{\,\,\,i}&=-\frac{i\hbar c}{2}\left(\overline{\Psi}\partial_i\Psi+\Psi\partial_i\overline{\Psi} \right)\label{eq:SC-0i}\\
            T^i_{\,\,\,j}&=\frac{\hbar^2}{2m}\left(\partial^i\overline{\Psi}\partial_j\Psi+\partial^i\Psi \partial_j\overline{\Psi} \right)-\delta^i_{\,j}\lag. \label{eq:SC-ij}
    \end{align}
    The momentum density is given by the $T^0_{\,\,\,j}$ components, where $j=1,2,3$ represents components along each spatial coordinate. By noting that $\partial \lag /\partial V =\overline{\Psi}\Psi$ is the probability density, the sourced continuity equation says
    \begin{equation}\label{eq:mom source}
        \partial_\mu T^\mu_{\,\,\,j}=\overline{\Psi}(-\partial_j V) \Psi,
    \end{equation}
     where it can be seen that the right-hand side represents a force expectation upon integration. For comparison, we put the left-hand side into integral form, simplify using the divergence theorem, and assume boundary terms vanish. This yields
     \begin{eqnarray}
         \int\!\! \partial_\mu T^\mu_{\,\,\,j} \,d^3x &=& \int\!\! \left(\frac{1}{c}\del{T^0_{\,\,\,j}}{t}+\del{T^i_{\,\,\,j}}{x^i} \right)d^3x \nonumber \\
         &=&\frac{d}{dt}\int\! \frac{1}{c}T^0_{\,\,\,j}\,d^3x \nonumber \\
         &=&-\frac{i\hbar}{2}\frac{d}{dt}\int\!\! \left( \overline{\Psi}\partial_j \Psi-\Psi\partial_j \overline{\Psi}\right)\,d^3x \nonumber\\
        &=&-\frac{i\hbar}{2}\frac{d}{dt}\int\!\! \big(2\overline{\Psi}\partial_j\Psi -\partial_j (\overline{\Psi}\Psi)\big)\,d^3x \nonumber\\
        &=&\frac{d}{dt}\int \!\overline{\Psi}\!\left(\frac{\hbar}{i}\partial_j\right)\!\Psi \,d^3x \nonumber\\
        &=&\frac{d\langle \hat{p}_j\rangle}{dt}, \label{eq:mom deriv}
     \end{eqnarray}
    where the momentum operator $\hat{p}_j\equiv \frac{\hbar}{i}\partial_j$ has been identified. We thus conclude that Eq. (\ref{eq:mom source}) in integral form is Ehrenfest's theorem for momentum,\cite{Takahashi1986}
    \begin{equation}\label{eq:ehrenfest momentum}
        \frac{d\langle\hat{\mathbf{p}}\rangle}{dt}=\langle -\nabla V\rangle.
    \end{equation}
    There are several points to make regarding this familiar result obtained by unfamiliar means. Locally, the negative gradient of an external potential---the external force---acts as a source of field momentum, which is captured by Eq. (\ref{eq:mom source}). Globally, this results in Ehrenfest's theorem for momentum. Moreover, this result does not require a probabilistic interpretation of the wavefunction, nor does it reference the field equations associated with Eq. (\ref{eq:shrodlag}). This indicates that Eq. (\ref{eq:source eq}), which makes no reference to a specific choice of Lagrangian, holds more generally for field theories with an external potential.  

    There is now a clear motivation to ask about the remaining Noether current representing energy density. The sourced continuity equation for this component is
    \begin{equation}\label{eq:nrg source}
        \partial_\mu T^\mu_{\,\,\,0}=-\frac{1}{c}\overline{\Psi}\del{V}{t}\Psi. 
    \end{equation}
    Putting the left-hand side in integral form as before, we now have an expression in terms of the Hamiltonian operator:
    \begin{eqnarray}
    \int\!\! \partial_\mu T^\mu_{\,\,\,0}\, d^3x &=& \int\!\! \left(\frac{1}{c}\del{T^0_{\,\,\,0}}{t} +\del{T^i_{\,\,\,0}}{x^i}\right)d^3x \nonumber \\
    &=& \frac{1}{c}\frac{d}{dt}\int \! T^0_{\,\,\,0}\,d^3x \nonumber \\
    &=& \frac{1}{c}\frac{d}{dt}\int\!\! \left(-\frac{\hbar^2}{2m}\partial^i\overline{\Psi}\partial_i\Psi -V\overline{\Psi}\Psi\right)d^3x \nonumber\\
    &=& \frac{1}{c}\frac{d}{dt}\int\!\left(\frac{\hbar^2}{2m}\overline{\Psi}\partial^i \partial_i\Psi -V\overline{\Psi}\Psi\right)d^3x \nonumber \\
    &=& -\frac{1}{c}\frac{d}{dt}\int\!\!\overline{\Psi}\!\left(-\frac{\hbar^2}{2m} \nabla^2 + V\right)\!\Psi\, d^3x \nonumber \\
    &=& -\frac{1}{c}\frac{d\langle \hat{H}\rangle}{dt}. \label{eq:ham deriv}
\end{eqnarray}
    Thus integrating both sides of Eq. (\ref{eq:nrg source}) and dividing out the factor of $-1/c$, we have an Ehrenfest-like relation for energy:
    \begin{equation}\label{eq:ehrenfest energy}
        \frac{d\langle \hat{H}\rangle}{dt} = \left\langle \del{V}{t}\right\rangle. 
    \end{equation}
    Just as the spatial derivatives of an external potential are sources of linear momentum, the time derivative of an external potential acts as a source of energy. Thus the expected energy is not a constant in time when a time-dependent potential is applied, but we still have a formal expression to track its rate of change. In Section \ref{moving infsqwell}, we demonstrate one example of this result holding true for an infinite square well with a moving floor.

    We now briefly cover the case of angular momentum, which yields a similar Ehrenfest relation. Consider a coordinate rotation 
    \begin{equation}
        x^\mu \to x^\mu +R^\mu_{\,\,\,\nu}x^\nu,
    \end{equation}
    where $R^\mu_{\,\,\,\nu}$ represents an infinitesimal rotation of the \textit{spatial} coordinates. We can show that there are six associated Noether currents 
    \begin{equation}\label{eq: angmom tensor}
        \mathcal{M}^{\mu\nu\sigma}\equiv x^\nu T^{\mu\sigma}-x^\sigma T^{\mu\nu},
    \end{equation}
    where $T^{\mu\nu}$ is the energy-momentum tensor as defined before with raised indices. A detailed derivation is provided in Appendix \ref{rotations}. We can use the Levi-Civita symbol to pick out the components representing the angular momentum density along each spatial coordinate,
    \begin{equation}
        \ell_i \equiv \frac{1}{c}\varepsilon_{ijk}x^jT^{0k}= \frac{1}{2c}\varepsilon_{ijk}\mathcal{M}^{0jk}.
    \end{equation}
    On one hand, we can now directly compute the divergence of Eq. (\ref{eq: angmom tensor}) with the product rule and pick out the terms of interest,
    \begin{equation}\label{eq:torque density}
        \frac{1}{2}\varepsilon_{ijk}\partial_\mu \mathcal{M}^{0jk} = \overline{\Psi} \tau_i \Psi,
    \end{equation}
    where $\tau_i\equiv \varepsilon_{ijk} x^i(-\partial^j V)$ is the $i$th component of the torque associated with $V$. It is thus clear that integration along all components of Eq. (\ref{eq:torque density}) yields the expected torque, $\langle \boldsymbol{\tau}\rangle$. On the other hand, computing the integral in a similar fashion to Eqs. (\ref{eq:mom deriv}) and (\ref{eq:ham deriv}) yields
    \begin{eqnarray}
        \frac{1}{2}\varepsilon_{ijk}\!\!\int\!\!\partial_\mu \mathcal{M}^{0jk}\,d^3x &=& \frac{1}{2}\varepsilon_{ijk}\!\! \int\!\!\left(\frac{1}{c}\del{\mathcal{M}^{0jk}}{t}+\del{\mathcal{M}^{\ell jk}}{x^\ell} \right)d^3x \nonumber\\
        &=& \frac{1}{2c}\varepsilon_{ijk}\frac{d}{dt}\!\int\!\! \mathcal{M}^{0jk}\,d^3x \nonumber \\
        &=& \frac{1}{2c}\varepsilon_{ijk}\frac{d}{dt}\!\int\!\!\left(x^j T^{0k}- x^k T^{0j} \right) d^3x \nonumber \\
        &=&\frac{d}{dt}\!\int\!\!\overline{\Psi}\left(\varepsilon_{ijk}\hat{x}^j\hat{p}^k\right)\Psi\,d^3x \nonumber \\
        &=& \frac{d\langle \hat{L}_i\rangle}{dt},
    \end{eqnarray}
    where $\hat{L}_i\equiv \varepsilon_{ijk}\hat{x}^j\hat{p}^k$ is the angular momentum operator. Comparing this with the integral of Eq. (\ref{eq:torque density}), we have Ehrenfest's theorem for angular momentum:\cite{Griffiths2018-vs}
    \begin{equation}\label{eq:ehrenfest ang mom}
        \frac{d\langle \hat{\mathbf{L}}\rangle}{dt}  =\langle \boldsymbol{\tau}\rangle.
    \end{equation}
    This derivation serves as an important template for deriving analogous results in the relativistic Klein-Gordon setting, which is the subject of Section \ref{KG}. 
    
    \subsection{Moving Floor Infinite Square Well}\label{moving infsqwell}
    We demonstrate one instance of the result in Eq. (\ref{eq:ehrenfest energy}) using the time-varying potential
\begin{equation}
    V(x,t)=\begin{cases}
        \Tilde{V}(t) & \text{if } 0<x<L\\
        \infty & \text{otherwise} 
    \end{cases}
\end{equation}
for some differentiable function $\Tilde{V}(t)$. This reduces to the more familiar stationary floor infinite square well when $\Tilde{V}(t)=0$. With boundary condition $\Psi(0,t)=\Psi(L,t)=0$, the full solution to the time-dependent Schr\"{o}dinger equation can be expanded into a complete orthonormal basis with time-varying coefficients
\begin{equation}
    \Psi(x,t)=\sum_{k=1}^\infty a_k(t)\sin\left( \frac{k\pi x}{L}\right).
\end{equation}
The coefficients $a_k(t)$ absorb the overall normalization and satisfy the residual requirement\cite{Franklin2013}
\begin{equation}
    0=\frac{\hbar^2k^2\pi^2}{2mL^2}a_k(t)-i\hbar \dot{a}_k(t)+a_k(t)\Tilde{V}(t).
\end{equation}
The solution to this set of differential equations is
\begin{equation}
    a_k(t)=A_k e^{-E_k t/\hbar}e^{-iw(t)/\hbar},
\end{equation}
where $E_k=\frac{\hbar^2k^2\pi^2}{2mL^2}$ are the energy states of the stationary square well and
\begin{equation}
    w(t)\equiv \int_{t'=0}^t \Tilde{V}(t')\,dt'
\end{equation}
is the additional phase picked up due to the time-dependent potential that does not depend on $k$. We can thus write the solution
\begin{equation}
    \Psi(x,t)=e^{-iw(t)/\hbar}\sum_{k=1}^\infty A_k e^{-iE_k t/\hbar}\sin\left( \frac{k\pi x}{L}\right),
\end{equation}
and the additional phase will disappear in any expression containing $|\Psi|^2$. With this, we can calculate
\begin{align}
    \frac{d\langle\hat{H}\rangle}{dt}  &=\frac{d}{dt}\!\int_0^L\! \overline{\Psi}\left(\frac{-\hbar^2}{2m}\del{^2}{x^2}+\Tilde{V}(t) \right)\Psi\,dx \nonumber\\
    &=\frac{-\hbar^2}{2m}\frac{d}{dt}\!\!\int_0^L\!\!\!\overline{\Psi}\del{^2\Psi}{x^2}\,dx+\frac{d}{dt}\!\!\int_0^L\!\!\! \Tilde{V}(t)|\Psi|^2dx.\label{eq:oscfloorham}
\end{align}
To handle the first integral, we use the orthonormal decomposition
\begin{eqnarray}
\int_0^L\!\!\overline{\Psi}\del{^2\Psi}{x^2}\,dx&=&\!\int_0^L\!\!\! -\delta_{jk} \!\left(\frac{k\pi}{L}\right)^2\overline{A}_jA_k e^{i(E_j-E_k)t/\hbar} \nonumber\\
&&\quad\quad \times \sin\left(\frac{j\pi x}{L} \right)\sin\left(\frac{k\pi x}{L} \right)dx\nonumber\\
&=&-\left(\frac{j\pi}{L}\right)^2|A_j|^2\!\!\int_0^L\!\!\sin^2\left(\frac{j\pi x}{L} \right)dx\nonumber\\
&=&-\left(\frac{j\pi}{L}\right)^2.
\end{eqnarray}
Since there is no time dependence here, the first term in Eq. (\ref{eq:oscfloorham}) vanishes. To handle the second term, we pass the time derivative into the integral and use the product rule:
\begin{eqnarray}
    \frac{d}{dt}\!\int_0^L\!\!\Tilde{V}(t)|\Psi|^2dx&=&\int_0^L\!\del{\Tilde{V}(t)}{t} |\Psi|^2dx+\Tilde{V}(t)\!\!\int_0^L\!\!\del{}{t}|\Psi|^2dx \nonumber\\
    &=&\int_0^L\! \overline{\Psi}\del{\Tilde{V}(t)}{t}\Psi\,dx \nonumber\\
    &=&\int_{-\infty}^\infty \!\overline{\Psi}\del{V(x,t)}{t}\Psi\,dx \nonumber\\
    &=&\left\langle \del{V(x,t)}{t}\right\rangle .
\end{eqnarray}
We have thus verified that Eq. (\ref{eq:ehrenfest energy}) holds for this system for any differentiable motion of the floor $\Tilde{V}(t)$.

\section{Complex Klein-Gordon Field Theory with Minimal Electric Potential Coupling}\label{KG}
The general outline of the previous section appears to be a way to generate Ehrenfest-like relations directly from a Lagrangian, without reference to the probabilistic interpretation of the wavefunction. We use this as motivation to develop analogous results in a relativistic setting using complex Klein-Gordon fields coupled with an electric potential. We use a minimal coupling given by the covariant derivative
\begin{equation}
    D_\mu \equiv \partial_\mu - i\frac{q}{\hbar} A_\mu, 
\end{equation}
where $q$ determines the coupling strength and
\begin{equation}
    A^\mu \equiv \begin{pmatrix}
        V/c\\
        \mathbf{A}
    \end{pmatrix}
\end{equation}
is the electromagnetic four-potential (note that $V$ now has a dimension of energy per unit charge). Since we are only interested in the scalar potential term, we set $\mathbf{A}=0$. In order to treat the potential as external to the theory, we will further omit the $\frac{1}{4\mu_0}F^{\alpha\beta}F_{\alpha\beta}$ term used in a scalar QED action. With these modifications, the Lagrangian we use is
\begin{align}
    \lag &= \overline{\left(D_\mu\phi\right)}\left(D^\mu\phi\right)+\frac{m^2c^2}{\hbar^2}\overline{\phi}\phi \nonumber \\
    &= \overline{\phi}_{,\mu}\phi_,^{\,\,\mu}+i\frac{qV}{\hbar c^2}\left(\overline{\phi}\Dot{\phi}-\phi\Dot{\overline{\phi}}\right)+\left(\frac{m^2c^2}{\hbar^2} -\frac{q^2V^2}{\hbar^2c^2} \right) \,\overline{\phi}\phi. \label{eq:em-kg-lagrangian}
\end{align}
Using the d'Alembertian operator $\square \equiv \partial_\mu \partial^\mu$, the field equations are 
\begin{eqnarray}
    &-\square\overline{\phi}+\left(\frac{m^2c^2}{\hbar^2} -\frac{q^2V^2}{\hbar^2c^2} \right)\overline{\phi}+2i\frac{qV}{\hbar c^2}\dot{\overline{\phi}}+\frac{iq\Dot{V}}{\hbar c^2}\overline{\phi}=0 \\
    &-\square\phi+\left(\frac{m^2c^2}{\hbar^2} -\frac{q^2V^2}{\hbar^2c^2} \right)\phi-2i\frac{qV}{\hbar c^2}\dot{\phi}-\frac{iq\Dot{V}}{\hbar c^2}\phi=0.
\end{eqnarray}
As before, we will first find the conservation law associated with a $U(1)$ symmetry, $\phi\to e^{i\theta}\phi$. The conserved current is
\begin{equation}\label{eq:J0}
    J^\mu=i\theta\left(\overline{\phi}\partial^\mu\phi-\phi\partial^\mu\overline{\phi}-2\frac{iq}{\hbar}A^\mu\overline{\phi}\phi \right).
\end{equation}
The 0-component can be simplified by separating the fields into their spatial and time-dependent components
\begin{equation}
    \phi = \psi(\mathbf{r})e^{-iEt/\hbar},
\end{equation}
and the complex conjugate. This allows us to write the time derivatives of the fields as
\begin{equation}\label{eq:phi dots}
    \dot{\phi}=\frac{E}{i\hbar}\phi \quad \text{and} \quad \dot{\overline{\phi}}=-\frac{E}{i\hbar}\overline{\phi}.
\end{equation}
After setting units, the conservation law reads
\begin{equation}
    \del{}{t}\rho+\nabla\cdot \mathbf{j}=0,
\end{equation}
where 
\begin{equation}\label{eq:rho}
    \rho\equiv \frac{2q}{\hbar^2c^2}(E-qV)\overline{\phi}\phi
\end{equation}
is interpreted as an electric charge density,\cite{Winter1959} and 
\begin{equation}
    \mathbf{j}\equiv i\frac{q}{\hbar}(\phi\nabla\overline{\phi}-\overline{\phi}\nabla\phi)
\end{equation}
is the corresponding electric current. Since electric charge conservation appears precisely where probability conservation appeared in the Schr\"{o}dinger field theory, we anticipate the same replacement to appear in the continuity equations associated with the energy-momentum tensor. Section \ref{rel. ehrenfest relations} shows that this is indeed the case.

\subsection{Relativistic Ehrenfest Relations} \label{rel. ehrenfest relations}
Given our Lagrangian that depends on $\phi$, its complex conjugate, their derivatives, and the electric potential $V$
\begin{equation}
    \lag=\lag(\phi,\overline{\phi},\partial_\mu\phi,\partial_\mu \overline{\phi},V),
\end{equation}
we can follow the exact same procedure outlined from Eq. (\ref{eq:spacetime translation}) to Eq. (\ref{eq:source eq}). Again defining the energy-momentum tensor as 
\begin{equation}\label{eq:KG ST}
    T^\mu_{\,\,\,\nu}\equiv \del{\lag}{\phi_{,\mu}}\phi_{,\nu}+\del{\lag}{\overline{\phi}_{,\mu}}\overline{\phi}_{,\nu}-\delta^\mu_{\,\,\,\nu}\lag,
\end{equation}
the continuity equation still reads as Eq. (\ref{eq:source eq}). However, the factor of $\partial \lag/\partial V$ in the source term is not as straightforward to calculate. Computing it explicitly gives
\begin{eqnarray}
    \del{\lag}{V}&=&\frac{iq}{\hbar c^2}\left(\overline{\phi}\dot{\phi}-\phi\dot{\overline{\phi}} \right)-2\frac{q^2V}{\hbar^2c^2}\overline{\phi}\phi \nonumber \\
    &=&\frac{iq}{\hbar c^2}\left(2\frac{E}{i\hbar}\overline{\phi}\phi\right)-2\frac{q^2V}{\hbar^2c^2}\overline{\phi}\phi \nonumber \\
    &=&\frac{2q}{\hbar^2c^2}(E-qV)\overline{\phi}\phi\nonumber\\
    &=&\rho,
\end{eqnarray}
where we used Eq. (\ref{eq:phi dots}) to simplify the time derivatives as before. Thus the derivatives of the potential are weighted by the electric charge density in the source term of Eq. (\ref{eq:source eq}), exactly taking the place of probability density in the Schr\"{o}dinger theory. The continuity equation is thus
\begin{equation}\label{eq:KG source eq}
    \partial_\mu T^\mu_{\,\,\,\nu}=-\rho\partial_\nu V,
\end{equation}
where the energy-momentum tensor components are 
\begin{align}
    T^0_{\,\,\,0}&=-\frac{1}{c^2}\Dot{\overline{\phi}}\Dot{\phi}-\nabla\overline{\phi}\cdot\nabla \phi -\left(\frac{m^2c^2}{\hbar^2}-\frac{q^2V^2}{\hbar^2c^2}\right)\overline{\phi}\phi \label{eq:KG-00}\\
    T^i_{\,\,\,0}&=\frac{1}{c}\left(\Dot{\overline{\phi}}\partial^i\phi+\Dot{\phi}\partial^i\overline{\phi} \right)\label{eq:KG-i0}\\
    T^0_{\,\,\,i}&=-\frac{1}{c}\left(\dot{\overline{\phi}}\partial_i\phi+\Dot{\phi}\partial_i\overline{\phi}\right)+i\frac{qV}{\hbar c}\Big(\overline{\phi}\partial_i\phi -\phi\partial_i\overline{\phi}\Big)\label{eq:KG-0i}\\ T^i_{\,\,\,j}&=\partial^i\overline{\phi}\partial_j\phi+\partial^i\phi\partial_j\overline{\phi}-\delta^i_{\,\,\,j}\lag. \label{eq:KG-ij}
\end{align}
The component $T^0_{\,\,\,0}$ is the Hamiltonian density or energy density (up to a sign) stored in the fields. Upon integration, there is no longer a familiar expression like the expected Hamiltonian in Eq. (\ref{eq:ham deriv}). Instead, we write the field Hamiltonian directly in terms of $T^0_{\,\,\,0}$:
\begin{equation}
    H\equiv \int -T^0_{\,\,\,0}\,d^3x.
\end{equation}
With this definition and assuming boundary terms vanish as before, the $\nu=0$ continuity equation in integral form reads
\begin{equation}
    \frac{d}{dt}H =\int\!\! \rho \del{V}{t}\,d^3x.
\end{equation}
Structurally, this is the same as Eq. (\ref{eq:ehrenfest energy}) with a charge density replacing probability density. In a similar fashion, we directly use the field momentum components
\begin{equation}
    P_i\equiv \int \frac{1}{c}T^0_{\,\,\,i}\,d^3x,
\end{equation}
and $\mathbf{P}=(P_1,P_2,P_3)$. Noting that $\mathbf{E}\equiv -\nabla V$ defines an electric field, the $\nu =1,2,3$ continuity equations yield
\begin{equation}
    \frac{d}{dt}\mathbf{P} = \int\!\! \rho \mathbf{E} \,d^3x.
\end{equation}
Stated in words, the rate of change of the field momentum is equal to the external Coulomb force acting on a charge distribution. This is a fitting counterpart to Ehrenfest's theorem, which relates the rate of change of expected momentum to the external force acting on a probability distribution. Of course, Ehrenfest's theorem for momentum is itself a quantum mechanical counterpart of Newton's second law, which relates the rate of change of mechanical momentum to the external force acting on a mass distribution. 

A similar result can be shown for angular momentum. We still have the six Noether currents associated with a spatial coordinate rotation
\begin{equation}\label{eq:KG angmom tensor}
        \mathcal{M}^{\mu\nu\sigma}\equiv x^\nu T^{\mu\sigma}-x^\sigma T^{\mu\nu},
\end{equation}
from which the angular momentum density components are
\begin{equation}
    \ell_i=\frac{1}{2c}\varepsilon_{ijk}\mathcal{M}^{0jk}.
\end{equation}
Now defining the field angular momentum
\begin{equation}
    L_i\equiv \int \ell_i \,d^3x,
\end{equation}
it is straightforward to show that
\begin{equation}
    \frac{1}{2}\varepsilon_{ijk}\int \partial_\mu \mathcal{M}^{\mu jk}\,d^3x = \frac{d}{dt}L_i, 
\end{equation}
assuming boundary terms vanish so that only the $\mu =0$ term survives integration. Compare this to an explicit calculation of the divergence of the four-current using the product rule, which after simplifying with Eq. (\ref{eq:KG source eq}) yields
\begin{equation}\label{eq:KG ang mom}
    \frac{1}{2}\varepsilon_{ijk}\partial_\mu \mathcal{M}^{\mu jk} =\varepsilon_{ijk}x^j(-\partial^k V) \rho,
\end{equation}
where the right-hand side is a torque per unit charge multiplied by the charge density $\rho$. The right-hand side is a torque per unit charge multiplied by the charge density $\rho$. Integrating and comparing with Eq. (\ref{eq:KG ang mom}) we have a relativistic Ehrenfest relation for angular momentum,
\begin{equation}
    \frac{d}{dt}\mathbf{L}=\int (\rho\mathbf{r}\times \mathbf{E})\,d^3x.
\end{equation}
Here $\rho\mathbf{r}=\rho(x^1,x^2,x^3)$ represents a dipole moment vector, and the electric field is defined as before. Since the integrand is weighted by a charge density, we say that the change in field angular momentum is equal to the external torque acting on a charge distribution.

\section{Conclusion}
This paper examined the effect of external scalar potentials on Lagrangian systems. When the potential is coordinate-dependent, such systems are not symmetric under a coordinate transformation and hence conservation of energy and momentum do not hold. By accounting for the response of the potential to spacetime translation, however, we find a continuity equation wherein the derivatives of the potential act as a source of field energy and field momentum. In integral form, these continuity equations express familiar Ehrenfest relations directly from a Schr\"{o}dinger Lagrangian. This approach makes no appeal to the probabilistic interpretation of the wavefunction, nor does it require explicit reference to the Schr\"{o}dinger equation. We thus postulate that Eq. (\ref{eq:source eq}) is a general way to derive Ehrenfest-like relations for a Lagrangian acted on by an external, coordinate-dependent scalar potential.

We then demonstrated one example of Ehrenfest-like relations for a complex Klein-Gordon Lagrangian coupled to an electric potential. These results for relativistic spin 0 particles have an electric charge density $\rho$ taking the place of the probability density $|\Psi|^2$ from the Schr\"{o}dinger Lagrangian. A natural progression of this work would be to replicate these results for a Dirac Lagrangian describing spin 1/2 particles.  

It should be noted that the derivation leading to Eq. (\ref{eq:source eq}) is not the only way to derive the Ehrenfest relations shown this paper, although other methods are far less elegant. One could instead explicitly calculate $\partial_\mu T^\mu_{\,\,\,\nu}$ using Eqs. (\ref{eq:SC-00}--\ref{eq:SC-ij}) for the Schr\"{o}dinger tensor and Eqs. (\ref{eq:KG-00}--\ref{eq:KG-ij}) for the Klein-Gordon tensor. This process is much more involved and requires extensive use of the field equations in order to arrive at the desired continuity equations. However, the correspondence of both methods lends support to the notion of treating a Lagrangian as an explicit function of an external scalar potential when one is applied. 

\appendix*   
\section{Noether Currents From Coordinate Transformations}
We discuss more generally the standard Noether currents spacetime translations and spatial rotations for a Lagrangian of two fields $\varphi$ and $\overline{\varphi}$. Generalizing to a Lagrangian with any number of fields is straightforward.

\subsection{Spacetime Translation} \label{spacetimetranslation}
    The infinitesimal spacetime translation $x^\mu\to x^\mu +a^\mu$ has the linearized field response 
    \begin{equation}
        \varphi(x^\mu+a^\mu)\approx \varphi(x^\mu)+a^\mu \partial_\mu \varphi,
    \end{equation}
    and the complex conjugate. After calculating the response of the Lagrangian (I) as a scalar function of the coordinates and (II) as a function of the fields and their derivatives (and simplifying with field equations), we see that the associated Noether current is
    \begin{equation}
        J^\mu=a^\nu\left(\del{\lag }{\varphi_{,\mu}}\varphi_{,\nu}+\del{\lag }{\overline{\varphi}_{,\mu}}\overline{\varphi}_{,\nu}-\delta^\mu_{\,\,\nu}\lag \right).
    \end{equation}
    In taking the divergence of this current, $\partial_\mu$ only acts on the expression inside the parentheses. This allows us to write the term inside as four Noether currents expressed as a mixed second-rank stress tensor, 
    \begin{equation}\label{eq:energy-momentum_tensor}
        T^\mu_{\,\,\,\nu}\equiv \del{\lag }{\varphi_{,\mu}}\varphi_{,\nu}+\del{\lag}{\overline{\varphi}_{,\mu}}\overline{\varphi}_{,\nu}-\delta^\mu_{\,\,\nu}\lag.
    \end{equation}
    We say that the four currents are associated with the energy and linear momentum densities stored in the fields. To see this explicitly, first note that 
    \begin{equation}
         T^0_{\,\,\,0} = \del{\lag}{\Dot{\varphi}}\Dot{\varphi}+\del{\lag}{\Dot{\overline{\varphi}}}\Dot{\overline{\varphi}} -\lag, 
    \end{equation}    
    and we define 
    \begin{equation}
        \pi(x^\mu)\equiv \del{\lag}{\dot{\varphi}} \quad \overline{\pi}(x^\mu)\equiv\del{\lag}{\dot{\overline{\varphi}}}
    \end{equation}
    to be the \textit{canonical} momentum densities conjugate to the fields $\varphi$ and $\overline{\varphi}$ respectively.\cite{Peskin:1995ev} We then see that this component of the stress tensor defines the Hamiltonian density $\mathcal{H}$ of the system,
    \begin{equation}\label{eq:hamdense}
        -\mathcal{H}\equiv \pi(x^\mu)\dot{\varphi}+\overline{\pi}(x^\mu)\dot{\overline{\varphi}}-\lag =T^0_{\,\,\,0}.
    \end{equation}
    Thus the Hamiltonian, or total energy of the system, is found by integrating the above expression over all three-space, 
    \begin{equation}
        H\equiv \int\!\! \mathcal{H}\, d^3x = -\int\!\! T^0_{\,\,\,0}\,d^3x.  
    \end{equation}
    Similarly, the spatial components give the \textit{physical} linear momentum stored in the field,
    \begin{equation}
        \mathcal{P}_i\equiv [\pi(x^\mu)\partial_i\varphi+\overline{\pi}(x^\mu)\partial_i\overline{\varphi}]=\frac{1}{c}T^0_{\,\,\,i},
    \end{equation}
    and
    \begin{equation}\label{eq:field-momentum}
        P_i\equiv \int\!\! \mathcal{P}_i \,d^3x=\int\!\!\frac{1}{c}T^0_{\,\,\,i}\,d^3x.
    \end{equation}
    
    \subsection{Spatial Rotation} \label{rotations}
    A rotation of spatial coordinates in classical mechanics is associated with the conservation of angular momentum. A rotation of spatial coordinates can be written in terms of a $4\times 4$ rotation matrix $A$ which transforms our position vector with $x^\mu\to Ax^\mu $ but leaves the time component unchanged. In the case of an infinitesimal transformation, the rotation matrix can be decomposed into
    \begin{equation}\label{eq:rotation-decomp}
        A\approx I+R,
    \end{equation}
    where $I$ is the identity and $R$ is an antisymmetric matrix with an infinitesimal parameter (omitted here for simplicity). We can represent this rotation as a mixed second-rank tensor $A^\mu_{\,\,\,\nu}$ applied to our position vector, and using the decomposition (\ref{eq:rotation-decomp}) we have
    \begin{equation}
        x^\mu \to x^\mu +R^\mu_{\,\,\,\nu}x^\nu.
    \end{equation}
    The field response to this transformation is approximated by Taylor expansion about $x^\mu$:
    \begin{equation}\label{eq:field rotation}
        \varphi(x^\mu+R^\mu_{\,\,\,\nu}x^\nu)\approx\varphi(x^\mu)+(\partial_\mu\varphi)R^\mu_{\,\,\,\nu}x^\nu,
    \end{equation}
    and complex conjugate. The Lagrangian, as a scalar function of coordinates, responds with
    \begin{equation}\label{eq:lagrot}
        \lag(x^\mu+R^\mu_{\,\,\,\nu}x^\nu)\approx \lag(x^\mu)+(\partial_\mu\lag)R^\mu_{\,\,\,\nu}x^\nu.
    \end{equation}
    To write the last term as an overall divergence, note that 
    \begin{align}
        \partial_\mu(\lag R^\mu_{\,\,\,\nu}x^\nu)&=(\partial_\mu \lag)R^\mu_{\,\,\,\nu}x^\nu+\lag R^\mu_{\,\,\,\nu}(\partial_\mu x^\nu)\nonumber\\
        &=(\partial_\mu \lag)R^\mu_{\,\,\,\nu}x^\nu+\lag R^\mu_{\,\,\,\nu}\delta^\nu_{\,\,\,\mu}\nonumber\\
        &=(\partial_\mu \lag)R^\mu_{\,\,\,\nu}x^\nu+\lag R^\mu_{\,\,\,\mu}\nonumber \\
        &=(\partial_\mu \lag)R^\mu_{\,\,\,\nu}x^\nu ,
    \end{align}
    where the last step holds since $R^\mu_{\,\,\,\nu}$ is antisymmetric and hence the diagonal components vanish. Thus we can write the change in the Lagrangian given in Eq. (\ref{eq:lagrot}) as an overall divergence,
    \begin{equation}
        \lag(x^\mu+R^\mu_{\,\,\,\nu}x^\nu)-\lag(x^\mu)=\partial_\mu(\lag R^\mu_{\,\,\,\nu}x^\nu).
    \end{equation}
    A similar calculation can be made for Eq. (\ref{eq:field rotation}). The Noether current for rotations is then
    \begin{equation}
        J^\mu=R^\sigma_{\,\,\,\nu}x^\nu\left(\del{\lag}{\varphi_{,\mu}}\varphi_{,\sigma}+\del{\lag}{\overline{\varphi}_{,\mu}} \overline{\varphi}_{,\sigma} -\delta^\mu_{\,\,\,\sigma}\lag \right).
    \end{equation}
    The term in parentheses is just the energy-momentum tensor $T^\mu_{\,\,\,\sigma}$ found in (\ref{eq:energy-momentum_tensor}). After moving the $\sigma$ index, this reduces to
    \begin{equation}
        J^\mu=R_{\sigma\nu} x^\nu T^{\mu\sigma}.
    \end{equation}
    We want to exploit the fact that $R_{\sigma\nu}$ is an antisymmetric tensor being multiplied into a third-rank contravariant tensor. First note that 
    \begin{equation}
        x^\nu T^{\mu\sigma}=\frac{1}{2}(x^\nu T^{\mu\sigma}+x^\sigma T^{\mu\nu})+\frac{1}{2}(x^\nu T^{\mu\sigma}-x^\sigma T^{\mu\nu}),
    \end{equation}
    where the first (second) term is symmetric (antisymmetric) under $\nu \leftrightarrow \sigma$ interchange. When we multiply the expression above by $R_{\sigma\nu}$  only the antisymmetric components survive the summation, so we have
    \begin{equation}
        J^\mu=\frac{1}{2}R_{\sigma\nu}(x^\nu T^{\mu\sigma}-x^\sigma T^{\mu\nu}).
    \end{equation}
    Discarding the constant term out front as before, we are left with six Noether currents
    \begin{equation}
        \mathcal{M}^{\mu\nu\sigma}\equiv x^\nu T^{\mu\sigma}-x^\sigma T^{\mu\nu}.
    \end{equation}
    To calculate the divergence, we use the product rule and find
    \begin{align}
        \mathcal{M}_{,\mu}^{\,\,\,\mu\nu\sigma} &= x^\nu\partial_\mu T^{\mu\sigma}-x^\sigma \partial_\mu T^{\mu\nu}+T^{\nu\sigma}-T^{\sigma\nu}\nonumber\\
        &=x^\nu\!\!\left(-\del{\lag}{V}\partial^\sigma V\right)-x^\sigma\!\! \left(-\del{\lag}{V}\partial^\nu V\right)+T^{\nu\sigma}-T^{\sigma\nu}, \label{eq:ang mom div}
    \end{align}
    where we have in general assumed that the Lagrangian depends on an external potential so that Eq. (\ref{eq:source eq}) holds. We have also not assumed a symmetric energy-momentum tensor, since that was the case for both systems studied in this paper. However, note that the linear momentum density is stored in the $T^{0k}$ components, so the $\mathcal{M}^{\mu jk}$ components will be relevant for studying angular momentum. The divergence of these terms is then 
    \begin{equation}\label{eq:general torque density}
        \mathcal{M}_{,\mu}^{\,\,\,\mu jk}=x^j\!\!\left(-\del{\lag}{V}\partial^k V\right)-x^k\!\! \left(-\del{\lag}{V}\partial^j V\right),
    \end{equation}
    where now $T^{jk}-T^{kj}=0$ since the spatial components of the energy-momentum tensor are in general symmetric. This result is used to write Eqs. (\ref{eq:torque density}) and (\ref{eq:KG ang mom}) using the Levi-Civita symbol. Thus conservation of angular momentum only holds when a spherically symmetric (if any) potential is applied. In general, however, the torque density in Eq. (\ref{eq:general torque density}) is a source of angular momentum from which Ehrenfest-like relations can be derived.

\begin{acknowledgments}
I would first like to thank Joel Franklin, my primary thesis advisor for this project at Reed College. Joel discovered a way to derive Ehrenfest's theorem for momentum by explicitly calculating the divergence of the energy-momentum tensor associated with the Schr\"{o}dinger Lagrangian, which inspired me to formulate the general expression in Eq. (\ref{eq:source eq}). Joel also came up with the choice of the relativistic Lagrangian Eq. (\ref{eq:em-kg-lagrangian}) among many other key suggestions in this paper. 

I would also like to thank Alexander Moll, my secondary thesis advisor, for providing extensive comments that led to this work.

\end{acknowledgments}

\bibliography{AJPTemplate}

\end{document}